\documentclass[%
 reprint,
superscriptaddress,
 amsmath,amssymb,
 aps,
 prl,
 onecolumn,
]{revtex4-2}

\usepackage{graphicx}% Include figure files
\usepackage{dcolumn}% Align table columns on decimal point
\usepackage{bm}% bold math
\usepackage{natbib}
\usepackage{hyperref}% add hypertext capabilities
\usepackage{siunitx}
\usepackage[caption=false]{subfig}
\usepackage{booktabs}
\usepackage[export]{adjustbox}
\usepackage{multirow}
\usepackage{physics}
\usepackage{xcolor}
\usepackage{accents}
\usepackage{color}

\usepackage{multibib}

\begin{document}

%\preprint{APS/123-QED}

% ALTERNATIVE TITLES
% 

\title{Feedback cooling Bose gases to quantum degeneracy}% Force line breaks with \\

\author{Matthew L. Goh}
 \email{matt.goh@merton.ox.ac.uk}%
 \affiliation{Department of Quantum Science and Technology, Research School of Physics, Australian National University, Canberra 2600, Australia}
 \affiliation{Department of Materials, University of Oxford, Parks Road, Oxford OX1 3PH, United Kingdom}
 \author{Zain Mehdi}
 \affiliation{Department of Quantum Science and Technology, Research School of Physics, Australian National University, Canberra 2600, Australia}%
\author{Richard L. Taylor}
 \affiliation{Department of Quantum Science and Technology, Research School of Physics, Australian National University, Canberra 2600, Australia}
\author{Ryan J. Thomas}
 \affiliation{Department of Quantum Science and Technology, Research School of Physics, Australian National University, Canberra 2600, Australia}%
 \author{Ashton S. Bradley}
 \affiliation{Department of Physics, Centre for Quantum Science, and Dodd-Walls Centre for
Photonic and Quantum Technologies, University of Otago, Dunedin, New Zealand}
\author{Michael R. Hush}
 \affiliation{Q-CTRL, Sydney, NSW Australia \& Los Angeles, CA USA}
\author{Joseph J. Hope}
 \affiliation{Department of Quantum Science and Technology, Research School of Physics, Australian National University, Canberra 2600, Australia}%
\author{Stuart S. Szigeti}
 \affiliation{Department of Quantum Science and Technology, Research School of Physics, Australian National University, Canberra 2600, Australia}%

\date{\today}% It is always \today, today,

% PACS, the Physics and Astronomy Classification Scheme.

\maketitle

\textbf{Degenerate quantum gases are instrumental in advancing many-body quantum physics~\cite{Gross2017} and underpin emerging precision sensing technologies~\cite{Szigeti2021}. All state-of-the-art experiments use evaporative cooling \cite{Anderson1995,Bradley1995,Davis1995} to achieve the ultracold temperatures needed for quantum degeneracy, yet evaporative cooling is extremely lossy: more than 99.9\% of the gas is discarded. Such final particle number limitations constrain imaging resolution, gas lifetime, and applications leveraging macroscopic quantum coherence. Here we show that atomic Bose gases can be cooled to quantum degeneracy using real-time feedback, an entirely new method that does not suffer the same limitations as evaporative cooling. Through novel quantum-field simulations and scaling arguments, we demonstrate that an initial low-condensate-fraction thermal Bose gas can be cooled to a high-purity Bose-Einstein condensate (BEC) by feedback control, with substantially lower atomic loss than evaporative cooling. Advantages of feedback cooling are found to be robust to imperfect detection, finite resolution of the control and measurement, time delay in the control loop, and spontaneous emission. Using feedback cooling to create degenerate sources with high coherence and low entropy enables new capabilities in precision measurement, atomtronics, and few- and many-body quantum physics \cite{Schreck2021}.} \par

Ultracold atomic samples are typically produced using cooling techniques largely unchanged from those used in the first realizations of atomic BECs in 1995~\cite{Anderson1995,Bradley1995,Davis1995}. Atomic samples are first brought to micro-Kelvin temperatures by laser cooling, followed by forced evaporation of the hotter atoms. Although this final evaporative stage gives the reduction to nano-Kelvin temperatures required for quantum degeneracy, it is an inherently lossy process that removes more than 99.9\% of atoms from the initial laser-cooled gas, even in highly-optimized scenarios~\cite{Wigley2016}. Although alternative cooling methods have been proposed and demonstrated, such as direct laser cooling~\cite{Stellmer2013} and sideband cooling~\cite{Urvoy2019}, they have thus far been limited to small atom-number samples of specific atomic species and also lack clear prospects for significantly increasing achievable atom flux. \par

Here we explore a novel and versatile technique for cooling ultracold gases to degeneracy using closed-loop feedback control, enabled by recent developments in optical control~\cite{Gauthier2016}. Feedback has emerged as a powerful quantum control solution for numerous quantum technologies and physical systems~\cite{Zhang2017}. This includes ultracold atoms, with proposed applications including improved atom laser stability~\cite{Wiseman2001,Thomsen2002,Haine2004,Johnsson2005}, the damping~\cite{Schemmer2017} and entangling~\cite{Wade2015, Wade2016} of a condensate's low-energy collective excitations, and the creation and stabilization of domain walls~\cite{Hurst2019} and magnetic phases~\cite{Hurst2020} in two-component BECs. In the control loop considered in this work, a dispersive optical measurement of the gas provides real-time information about the atomic gas density, which is fed back to an optical potential controller and used to damp density fluctuations in the gas. By feeding back information on the atomic density evolution, this feedback potential rapidly extracts energy from the atomic cloud without removing atoms, with atom losses arising only through spontaneous emission caused by the optical imaging. Thus, the trade-off between cooling efficiency and atom loss is fundamentally different to that of evaporative cooling. We demonstrate that this control scheme is capable of cooling a thermal, low-condensate-fraction Bose gas to degeneracy under realistic experimental conditions, building upon previous work demonstrating the control of coherent spatial excitations in a zero-temperature BEC~\cite{Szigeti2009,Szigeti2010,Hush2013}.\par

%%%%%%%FIGURE 1%%%%%%%%%%
\begin{figure}
\includegraphics[width=\textwidth]{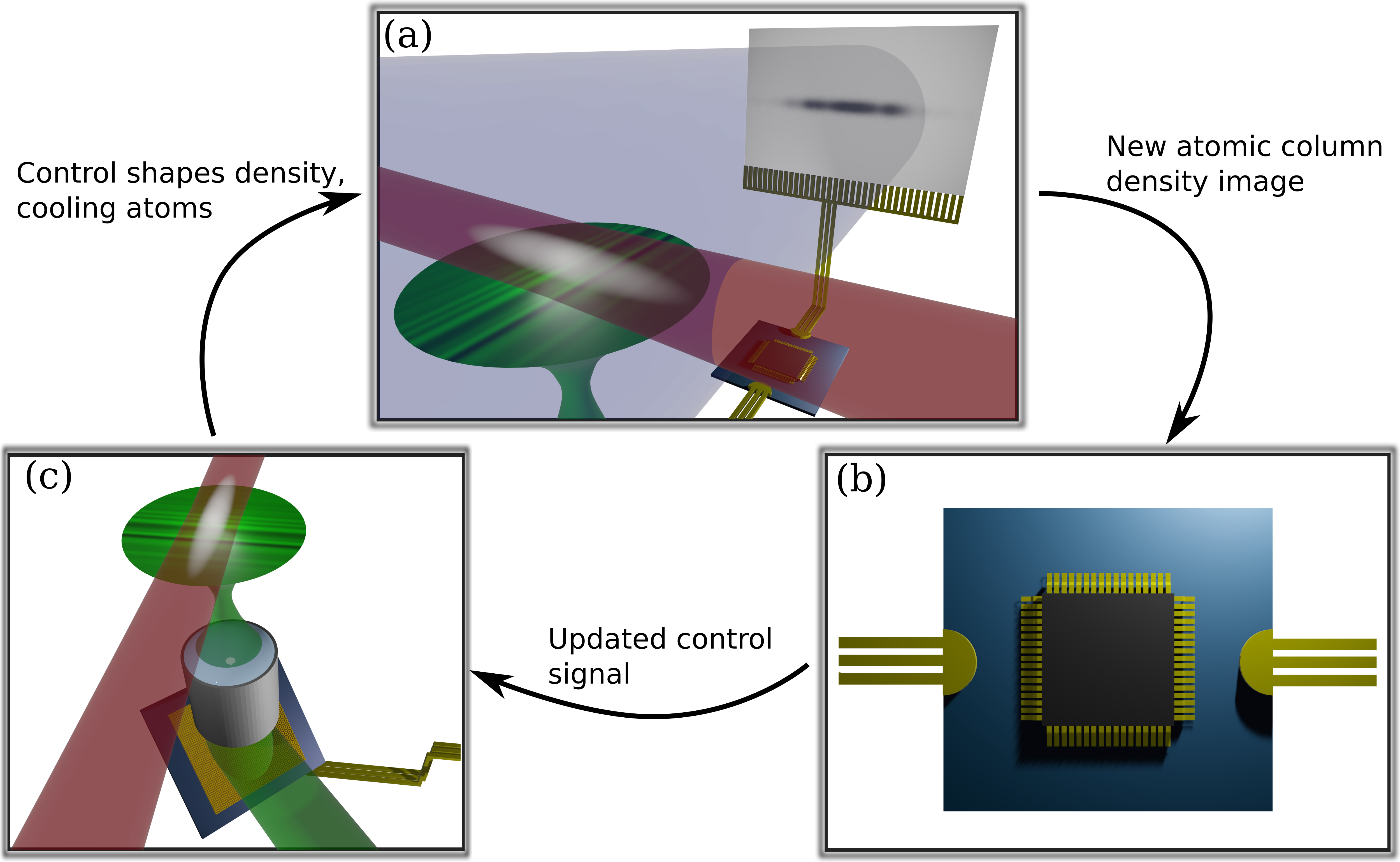}
\caption{\label{fig:systemsetup} \textbf{Feedback cooling process.} \textbf{(a)} A quasi-1D Bose gas (grey ellipse) is formed by trapping at the focus of a far red-detuned laser beam (red). An image is non-destructively captured by illuminating the atoms with a weak, off-resonant laser beam (blue), which creates a measurement signal proportional to the atomic column density. \textbf{(b)} A processor, such as a field-programmable gate array, microcontroller, or computer, reads and stores the image data and calculates a control signal from current and past atomic density information. \textbf{(c)} Through a spatial light modulator (SLM), such as a digital micro-mirror device (blue and gold surface), the control signal shapes the intensity of a far off-resonant laser beam (green) such that its image on the atoms is proportional to the time-derivative of the atomic column density in the elongated direction (bright and dark bands). Thus, real-time atomic density information is fed back via the control signal and used to remove density fluctuations in the gas, thereby removing energy and cooling the atoms. }
\end{figure}
%%%%%%%%%%%%%%%%%%%%%%

In detail, we model the experimental proposal depicted in Fig.~\ref{fig:systemsetup}, which is entirely within the constraints of demonstrated experimental capabilities. We consider a gas of ${}^{87}\text{Rb}$ atoms prepared in $\ket{F=1,m_F=1}$ and confined in a prolate harmonic trap with potential $V_0(x,y,z)=m\left(\omega_x^2 x^2 + \omega_\perp^2 y^2 + \omega_\perp^2 z^2\right)/2$, where the axial confinement is much tighter than the longitudinal confinement ($\omega_\perp \gg \omega_x$). This geometry, which is readily realized in experiments using magnetic or optical fields~\cite{Grlitz2001,Meyrath2005}, strongly suppresses density fluctuations in the $y$ and $z$ dimensions, admitting an effective-1D description (Methods). Real-time information about the atomic density is obtained using non-destructive spatial imaging~\cite{Andrews1996,Bradley1997}. By illuminating the atomic sample along the tightly-confined $z$ dimension by a laser far-detuned from the $5^2{\rm S}_{1/2}\to 5^2{\rm P}_{3/2}$ transition, the column density of the atomic cloud is imprinted on the phase of the light, which is continuously measured with a CCD camera. This provides a continuous, non-destructive weak measurement of the atomic cloud's spatial density.

Information about the atomic density evolution is fed back by manipulating the atomic cloud using a high-resolution optical potential $V_{\rm C}(x,t)$ controlled by a spatial light modulator (SLM), such as that of Ref.~\cite{Gauthier2016}. High-speed configurable optical potentials of this kind are a recent innovation, enabling controls superior to those previously considered in Refs.~\cite{Haine2004,Szigeti2009,Szigeti2010,Hush2013}. Our control is inspired by canonical reservoir interactions in the semiclassical theory of high-temperature Bose gases, which depend upon the atomic density current and provide highly efficient, coherent damping of density fluctuations~\cite{Blakie2008, McDonald2020}. Specifically, the derivative signal of the measurement is spatially filtered and used to construct the control (Methods), resulting in an effective viscous damping force $\bm{F}\propto-\bm{v}$ on the atoms. This cools the atomic cloud up to a limit predominantly determined by the measurement's signal-to-noise ratio (SNR).\par

It is highly non-trivial to faithfully yet tractably simulate the feedback cooling of a continuously-monitored Bose gas from high temperatures (i.e. near the critical temperature), where thermal fluctuations dominate, to near-degeneracy where quantum fluctuations dominate. We had to develop specialist multimode quantum-field theoretic techniques precisely for this task (Methods). A realistic full-field quantum model of the atomic Bose gas under the continuous optical imaging and feedback described above is given by the (It\^{o}) conditional master equation~\cite{Szigeti2009}
\begin{equation}
    d\hat{\rho}_c=\underbrace{-\frac{i}{\hbar}\left[\hat{H},\hat{\rho}_c\right]dt}_{\text{Trapping, control \& scattering}}+\underbrace{\alpha\int dx \mathcal{D}\left[\hat{M}_r(x)\right]\hat{\rho}_c dt}_{\text{Decoherence}} + \underbrace{\sqrt{\eta\alpha}\int dx \mathcal{H}\left[\hat{M}_r(x)\right]\hat{\rho}_c dW(x,t)}_{\text{Measurement innovations}},
    \label{eqn:conditionalmasterequation}
\end{equation}
where $\hat{\rho}_c$ is the conditional density operator for the atomic ensemble, capable of describing the real-time dynamics of an individual experimental realisation (with a specific measurement record). The Hamiltonian
\begin{equation}\hat{H}\equiv\int dx \hat{\Psi}^\dagger(x)\left(-\frac{\hbar^2}{2m}\frac{\partial^2}{\partial x^2 }+\frac{1}{2}m\omega_x^2 x^2+V_{\rm C}(x,t)\right)\hat{\Psi}(x)+\frac{U_{\rm 1D}}{2}\int dx \hat{\Psi}^\dagger(x)\hat{\Psi}^\dagger(x)\hat{\Psi}(x)\hat{\Psi}(x)
\end{equation}
describes the unitary evolution of the atoms under trapping, optical control and interatomic scattering, where $\hat{\Psi}(x)$ is the atomic field operator satisfying $[\hat{\Psi}(x), \hat{\Psi}^\dag(x)] = \delta(x - x')$. The dispersive effect of the imaging laser is described by the measurement operator $\hat{M}(x)\equiv \int dx' \hat{\Psi}^\dagger(x')K(x-x')\hat{\Psi}(x')$, where the effect of optical diffraction is captured by the point-spread kernel $K(x)$ (Methods). The measurement acts through the superoperators $\mathcal{D}[\hat{L}]\hat{\rho}\equiv \hat{L}\hat{\rho}\hat{L}^\dagger-\frac{1}{2}(\hat{L}^\dagger\hat{L}\hat{\rho}+\hat{\rho}\hat{L}^\dagger\hat{L})$, describing decoherence, and $\hat{\mathcal{H}}[\hat{L}]\hat{\rho}\equiv\hat{L}\hat{\rho}+\hat{\rho}\hat{L}^\dagger-\langle\hat{L}+\hat{L}^\dagger \rangle\hat{\rho}$, describing updates to the quantum state based on probabilistic observations from quantum measurement (innovations), with $dW(x,t)$ a Wiener increment satisfying $dW(x,t) dW(x',t) = \delta(x-x') dt$. The measurement is parameterised by a quantum efficiency $\eta\in (0,1]$ and an effective measurement strength $\alpha$, which scales with laser intensity and may be treated as a free parameter (further details in Methods). The feedback control potential $V_{\rm C}(x,t)$ is directly computed from the measurement signal, and has a SNR inversely proportional to $\sqrt{\eta\alpha}$. It therefore contains information about the atomic density dynamics and a measurement-induced backaction noise. To model the limited spatial resolution $r_c$ of the control, $V_{\rm C}$ is convolved with a point-spread Gaussian kernel of FWHM $r_c$. A time lag $\tau$ within the control loop is readily modelled by temporally translating $V_{\rm C}(x,t)\to V_{\rm C}(x,t-\tau)$.\par

The full-field model in Eq.~(\ref{eqn:conditionalmasterequation}) requires an intractably large many-body Hilbert space and cannot be exactly solved for more than a few modes and small numbers of atoms. Furthermore, standard approximate techniques used in ultracold atomic simulations are also unsuitable for solving Eq.~(\ref{eqn:conditionalmasterequation}) for our problem of interest. Traditional mean-field methods are only suitable for zero-temperature systems and fail to capture critical quantum correlations induced by the measurement~\cite{Hush2013}. Linearised treatments such as Bogoliubov theory can only model near-equilibrium collective excitations well below the critical temperature in already highly-degenerate scenarios. Finite-temperature classical-field methods and related phase-space methods (such as the stochastic projected Gross-Pitaevskii equation and the truncated Wigner method, respectively \cite{Blakie2008}) cannot faithfully reproduce the conditional dynamics induced by the measurement. The number-phase Wigner (NPW) particle filter~\cite{Hush2013}, a full-field phase-space method which faithfully represents the measurement, cannot scalably represent high-temperature thermal states in its existing form, but forms the foundation of our novel approach. In Methods, we describe how projective methods can be used to scalably and self-consistently extend NPW to finite-temperature states with a majority of non-condensed thermal atoms. These are crucial numerical techniques that enable us, for the first time, to perform dynamical simulations that include the multimode nature of the quantum gas and correctly account for the conditional dynamics of an atomic ensemble under dispersive monitoring across the transition from thermal to highly-degenerate regimes.

%%%%%%%%%FIGURE 2%%%%%%
\begin{figure}
\includegraphics[width=\textwidth]{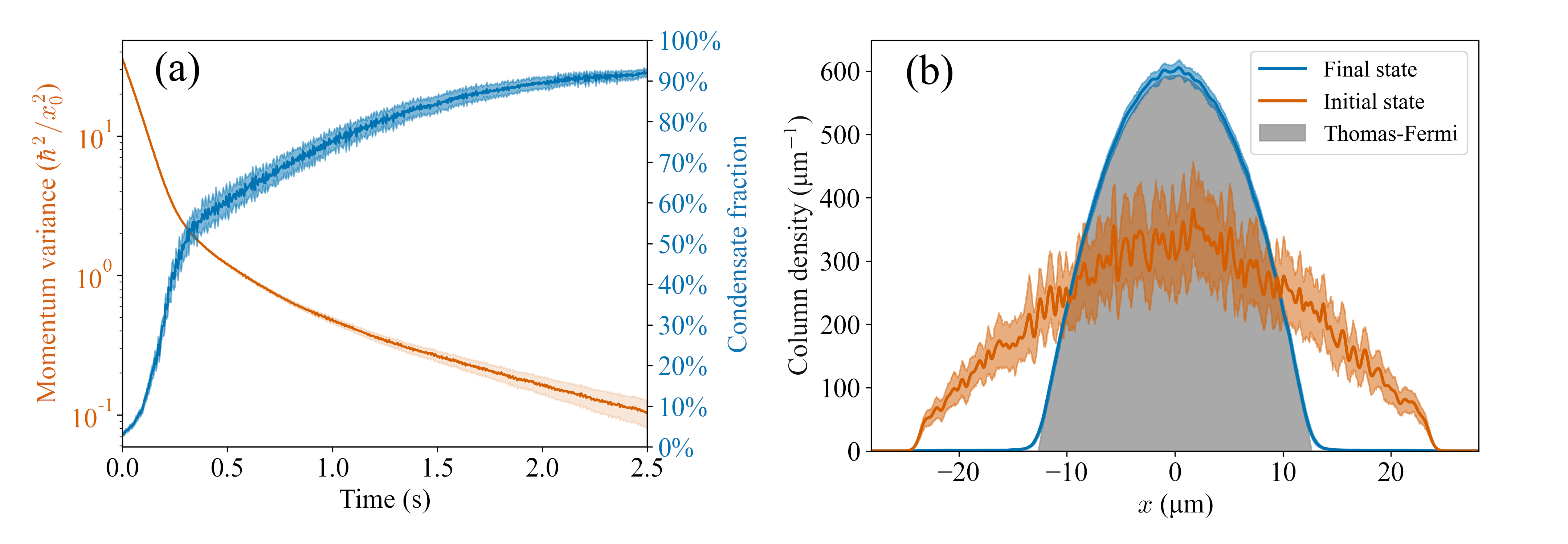}
\caption{\label{fig:dynamics} \textbf{Results of the feedback cooling process.} \textbf{(a)} The change in condensate fraction and momentum variance during feedback cooling of a cloud of $10^4$ ${}^{87}\text{Rb}$ atoms ($\omega_x=2\pi\times 40$ Hz). Over $2.5$s of cooling, the condensate fraction is increased from $(3.5\pm0.2)\%$ to $(92.1\pm0.5)\%$, and the momentum variance is reduced by a factor of $320$, key signatures that a highly pure BEC has been created. Momentum variance is given in units of $\hbar^2/x_0^2$, where $x_0=\sqrt{\hbar/m\omega_x}$ is the natural longitudinal length scale of the prolate trap. \textbf{(b)} Column densities of the initial thermal cloud ($3.5$\% condensate), feedback-cooled BEC ($92.1$\% condensate), and the analytic column density for a pure BEC in the Thomas-Fermi ground state. The column density profile of the condensate at the end of the cooling matches the Thomas-Fermi profile extremely well, as expected for a Bose gas that is highly pure and close to its ground state. Transparent shaded regions in both subfigures represent bootstrapped $95$\% confidence intervals.}
\end{figure}
%%%%%%%%%%%%%%%%%%%%%
Our simulations show that a Bose gas can be efficiently cooled from a mostly thermal cloud to a highly pure BEC. Figure \ref{fig:dynamics} demonstrates the dynamics of feedback cooling for a cloud of $10^4$ atoms. Over $2.5$ seconds of cooling, the condensate fraction is increased from $(3.5\pm0.2)\%$ to $(92.1\pm0.5)\%$. The momentum variance is reduced by a factor of $320$ and the spatial density of the final state closely approximates the Thomas-Fermi profile, both of which are expected for a pure, zero-temperature BEC. Supplementary Video 1 shows the dynamics of the feedback cooling process, and provides intuition for the relevant timescales, length scales, and operating principles. These results show that under ideal conditions, closed-loop feedback control can produce a high-purity BEC from an initial thermal sample.

%%%%%%%%%%%%%% FIGURE 3 %%%%%%%%%%%%%%%%
\begin{figure}
\includegraphics[width=0.5\textwidth]{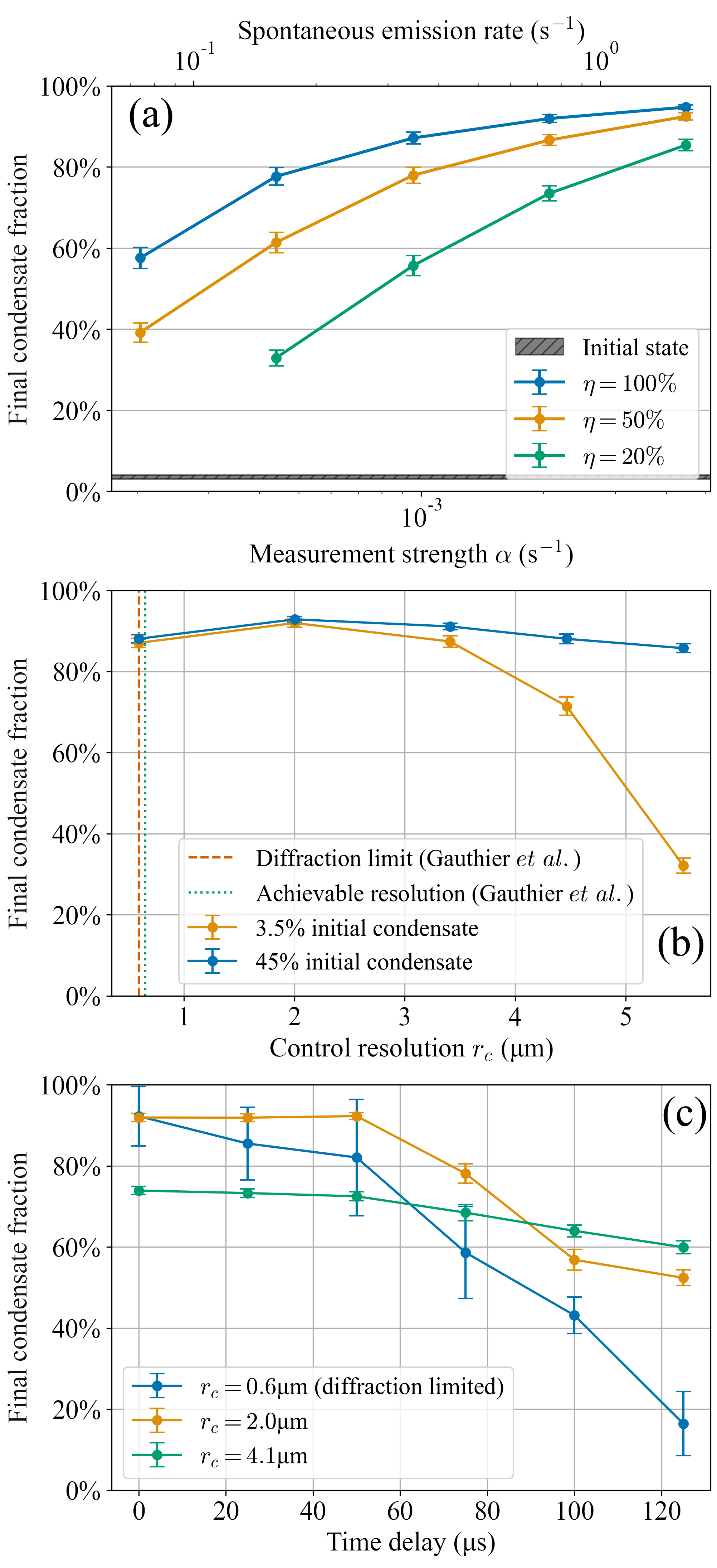}
\caption{\label{fig:parametervariations} \textbf{The effect of experimental imperfections on feedback cooling.} \textbf{(a)} The dependence of final condensate fraction on measurement strength $\alpha$ and quantum efficiency $\eta$. A higher measurement strength results in superior cooling due to improved SNR in the feedback loop. Lower quantum efficiency reduces the SNR and thus degrades the cooling, but can be compensated by increasing $\alpha$. Efficient cooling is still possible at quantum efficiencies as low as $20\%$, indicating feedback cooling is sufficiently robust to imperfect detection efficiency to be practical. \textbf{(b)} The dependence of final condensate fraction on the control resolution set by the SLM, compared for initial condensate fractions of $(3.5\pm0.2)\%$ and $(45\pm2)\%$. We also note the optical diffraction limit ($\lambda=532\text{nm}$, 0.45 NA) and achievable control resolution for the configurable optical potential demonstrated by Gauthier \textit{et al.} \cite{Gauthier2016}. The cooling is highly robust to imperfections in control resolution, with uncompromised performance for resolutions up to six times larger than the diffraction limit at these initial temperatures. For the hotter initial state, a smaller control resolution threshold is required for effective cooling due to the presence of higher-frequency spatial excitations. \textbf{(c)} The dependence of final condensate fraction on a time delay in the feedback loop at three difference control resolutions. At all tested resolutions, the cooling is robust up to a $50\mu\text{s}$ lag time, which is likely realizable in experiment. The effect of a time delay is magnified at smaller control resolutions, since the feedback's higher-frequency components act on a timescale shorter than the delay. Thus, the lag time should be considered when choosing the optimal control resolution for an experiment. Error bars on all subfigures represent bootstrapped 95\% confidence intervals.}
\end{figure}
%%%%%%%%%%%%%%%%%%%%%%%%%%%%%%%%%%%%%%%%%
We now quantify the impact of key limitations of a realistic experiment on effectiveness of cooling, and demonstrate our feedback control scheme to be robust to these effects within experimentally achievable parameters. Figure~\ref{fig:parametervariations} demonstrates the dependence of the condensate fraction after 2.5 seconds of cooling on measurement strength, quantum efficiency, spatial resolution of the control, and time delay in the feedback loop. 

Figure~\ref{fig:parametervariations}(a) shows that the final condensate fraction is reduced for lower quantum efficiencies, as the measurement SNR is degraded. Note that within our model inefficient detection and additional technical noise in the measurement signal (e.g. electronic noise) equate to a lowered quantum efficiency. Notably, the final condensate fraction strictly increases with measurement strength in this regime, indicating that the cooling performance is \textit{information-limited} - i.e. the improvement in cooling rate from larger measurement strength always exceeds the increase in heating from measurement backaction. Thus, the poorer cooling performance at lower quantum efficiencies can be overcome by increasing $\alpha$ (either by increasing laser intensity or decreasing detuning). Although this increases atom loss from spontaneous emission, the total atom loss is competitive with state-of-the-art evaporative cooling for $10^4$ atoms, and will be significantly reduced for larger atomic ensembles, as discussed below. \par

Figure~\ref{fig:parametervariations}(b) shows the effects of the control spatial resolution $r_c$ (the length scale of the control potential constructed by the SLM). The cooling process is highly robust to imperfect control resolution, with uncompromised cooling for resolutions up to a factor of six worse than the diffraction limit at the tested initial temperatures, and an optimal control resolution of $r_c\approxeq 2\,\mu$m that is well within current capabilities. The effect appears to be temperature-dependent, with hotter initial states requiring smaller control resolutions due to the increased spatial frequency of thermal excitations at higher temperature. Since spatial resolutions within $5\%$ of the optical diffraction limit are readily achievable in SLMs with current digital micromirror devices \cite{Gauthier2016}, it is likely that a BEC can be created via feedback from thermal clouds even hotter than those simulated here. Although this spatial resolution prevents the control from directly acting on the highest-frequency thermal excitations, interatomic scattering couples these uncontrolled modes to the lower-energy controllable modes, enabling the indirect cooling of excitations with wavelengths shorter than $r_c$~\cite{Haine2004,Szigeti2010}. This mechanism can be seen in Supplementary Video 1 and allows a low-resolution control to cool the high-frequency excitations of high temperature clouds. \par

Finally, in Fig.~\ref{fig:parametervariations} (c) we demonstrate robustness to time lag in the feedback loop, which may originate from the camera shutter speed, the filtering process, or the SLM. With current technology, the feedback loop would be bottlenecked by the SLM, for which a $20$kHz switching speed has been demonstrated~\cite{Gauthier2016}, setting a best-case lag of $50\,\mu$s. Fortunately, for the optimal control resolution of $r_c\approx2.0\,\mu$m, the cooling performance is unaffected by time delays up to $50\,\mu$s, and degrades beyond that point. This robustness is partially due to the relatively large value of $r_c$ compared to the smallest achievable resolution, which provides low-pass filtering of the atomic density signal and thereby excludes high-frequency components that would make the control sensitive to very small time delays. This is clear from the results using a smaller control resolution $r_c=0.59\,\mu$m, for which feedback lag degrades the performance more severely and at shorter lag times. Conversely, larger control resolutions increase robustness to larger time delays, at the expense of lower overall cooling performance at shorter time lags. This allows lag times larger than $50\,\mu$s to be accommodated; for example, $r_c=4.1\,\mu$m is significantly more robust to larger lag times than $r_c=2.0\,\mu$m, with only small changes in cooling performance up to a more substantial lag of $\tau=125\,\mu$s. \par

\begin{figure}
\includegraphics[width=0.47\textwidth]{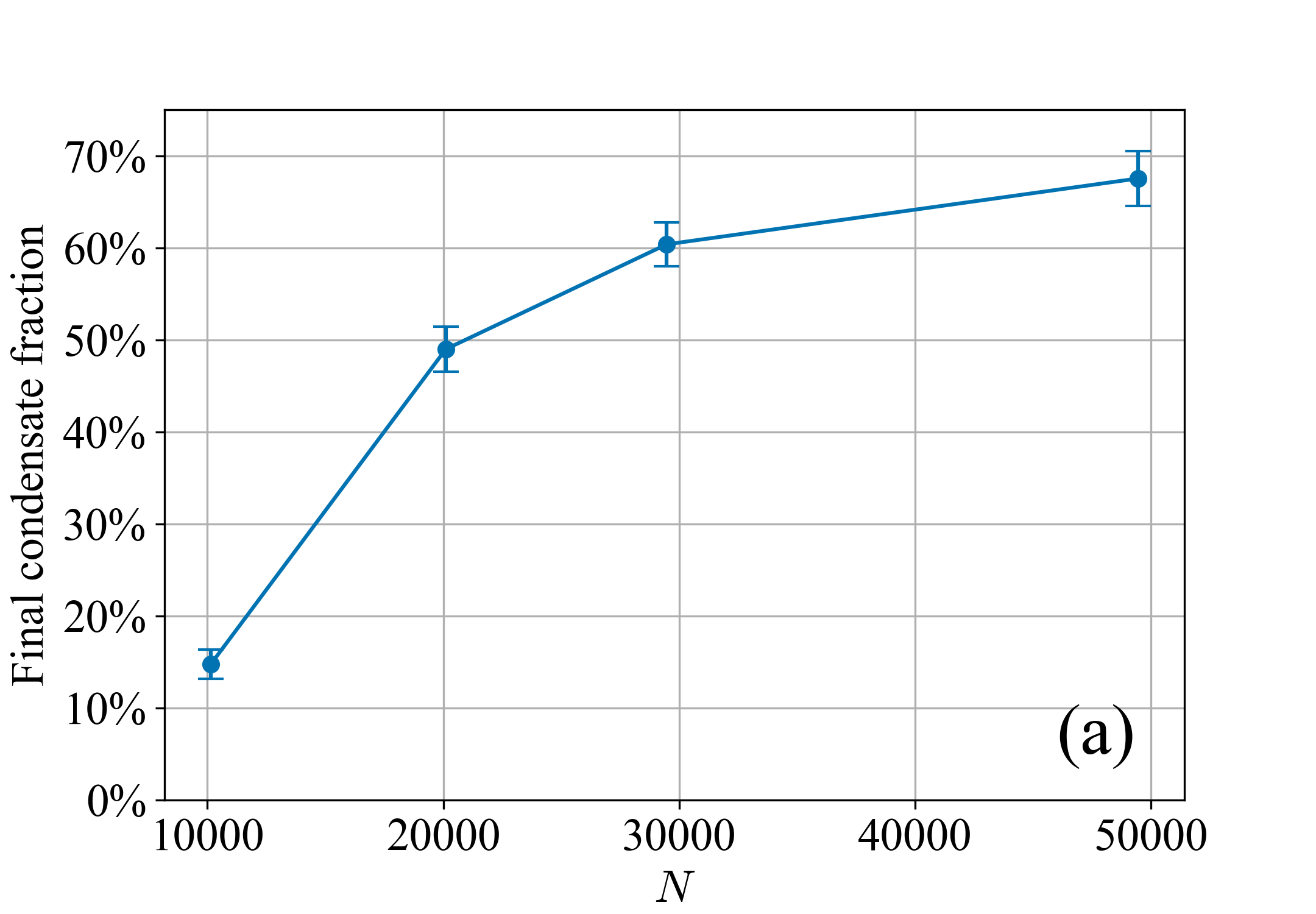}
\quad
\includegraphics[width=0.47\textwidth]{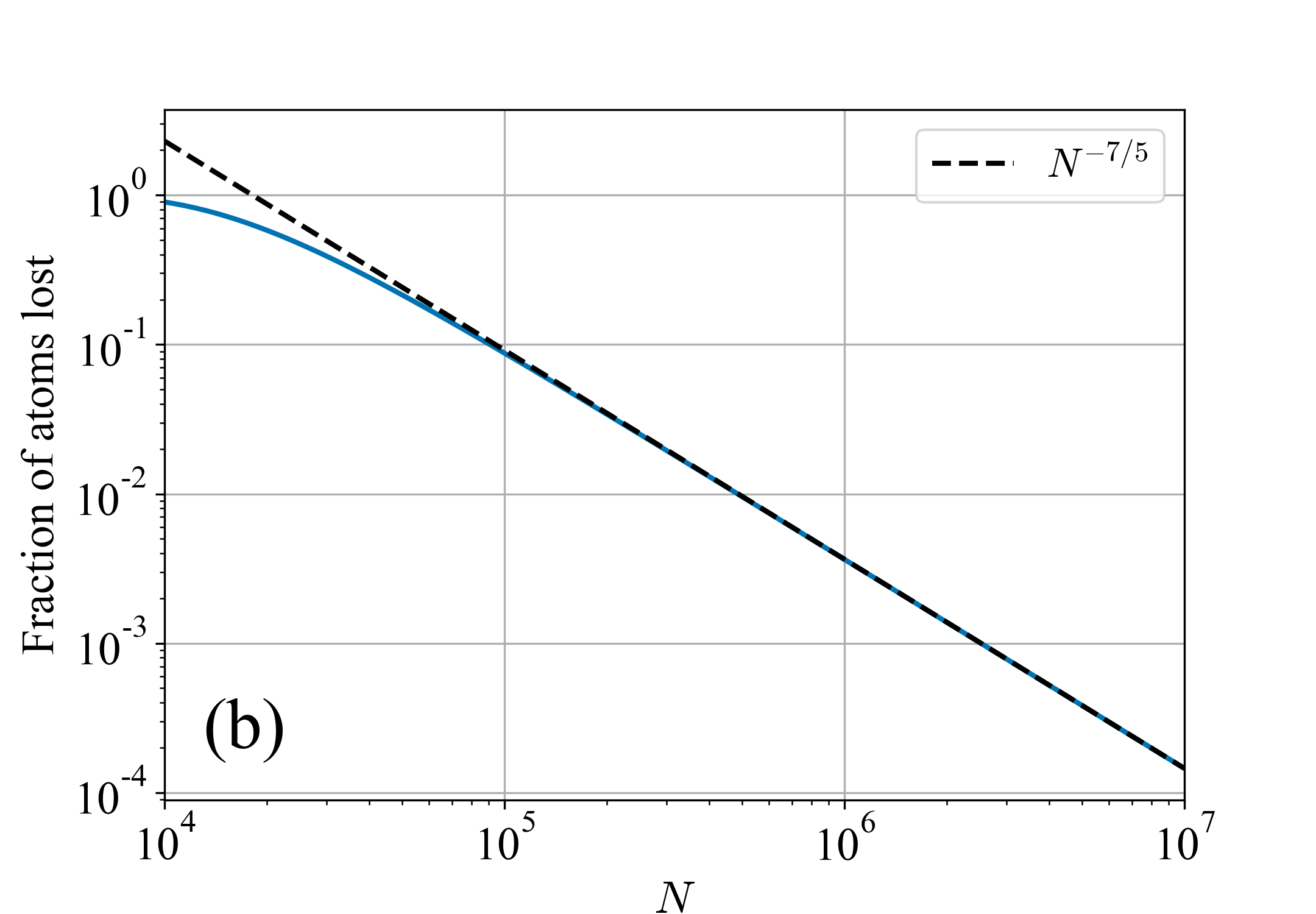}
\caption{\label{fig:atomloss}\textbf{Projected scaling of feedback cooling efficiency technique with atom count.} \textbf{(a)} The dependence of final condensate fraction on number of atoms $N$ in a highly information-limited regime. The detuning and intensity of the imaging laser are fixed such that only $5$\% of atoms are lost to spontaneous emission in each simulation. At a fixed loss rate, the cooling improves significantly with larger numbers of atoms, due to the increase in measurement SNR at higher optical density. Simulations beyond $5\times10^4$ atoms were not possible due to computational requirements exceeding the capabilities of available supercomputer facilities. Error bars represent bootstrapped $95$\% confidence intervals. \textbf{(b)} Fraction of atoms lost in cooling as a function of initial number of atoms $N$ for fixed SNR (assuming shot-noise limited measurement), corresponding to $90\%$ loss for $N=10^4$ atoms. Here we fix $\eta=1$ and $r_c= 2\mu\text{m}$.}
\end{figure}

The proposed feedback cooling scheme has a fundamentally different mechanism of atomic loss compared to evaporative cooling; namely spontaneous emission induced by the dispersive imaging. This loss is easily estimated from the parameters of our model (Methods), with the loss rate directly proportional to the measurement strength $\alpha$. This implies a tradeoff between this fundamental loss channel and the SNR of the measurement. This is particularly important for the scheme considered in this work,  as the control is constructed directly from the most recent measurement signal and thus the SNR strongly limits the final condensate fraction and number in the parameter regimes we consider. For the simulation results presented above, there would be significant atomic losses due to spontaneous emission - for the measurement strength of $\alpha\approx 10^{-3}$ considered in Figure~\ref{fig:dynamics}, approximately $90\%$ of atoms are lost over the $2.5$s of cooling. However, this significant loss is not a fundamental limitation of our scheme; due to the computational complexity of our simulation method, it was not tractable to simulate atom numbers significantly larger than $N=10^4$. The measurement SNR is impoverished for such small atom number clouds as they have relatively low optical densities, which must be compensated for with relatively strong measurement strengths in order to achieve efficient feedback cooling. \par 

However, for larger initial atom numbers, the increased optical density  of the cloud allows effective feedback cooling at significantly weaker measurement strengths, thus suppressing atom loss due to spontaneous emission. This is evident in our simulations - Fig.~\ref{fig:atomloss}(a) demonstrates that even a modest factor of $5$ increase of the initial atom number dramatically improves the cooling process for a weak measurement chosen to give a fixed atomic loss of $5\%$ over $2.5$s of cooling. Furthermore, we extend our analysis to much larger atomic ensembles than can be tractably simulated by considering the atom-number dependency of the measured density SNR for an interacting Bose gas. As the SNR of a continuous measurement signal is ill-defined, we instead consider a related quantity which we term the \emph{time-integrated SNR} that quantifies the quality of the measurement signal over the entire cooling process as a function of the trapping geometry, pixel size, detector efficiency, spontaneous emission rate, and atom number (see Methods). In Fig.~\ref{fig:atomloss}(b) we demonstrate that, for a fixed measurement SNR, the fraction of atoms lost due to spontaneous emission decreases with atom number as $N^{-7/5}$. This implies that the same steady-state condensate fraction demonstrated in Fig.~\ref{fig:dynamics} with roughly $90\%$ atomic loss for $N=10^4$ initial atoms can in principle be achieved with less than $1\%$ atomic loss for an initial sample of $N=5\times10^5$ atoms, provided the efficacy of feedback cooling remains limited chiefly by the SNR of the measurement signal. These results suggest that the creation of feedback-cooled BECs scale favourably with larger numbers of atoms, allowing low-loss cooling of very large atomic ensembles to degeneracy. We therefore anticipate that closed-loop feedback cooling may surpass current limitations on BEC atom count imposed by evaporative cooling.

Further innovations can likely be made. As a proof of principle, we focused here on atomic clouds where the dynamics are largely restricted to a single trapping dimension. This was due to computational constraints rather than any practical experimental limitation. Indeed, feedback control may be a much more effective cooling technique for quasi-2D atomic clouds, which have much faster thermalisation rates than 1D systems and therefore would give more efficient coupling between the high-frequency modes (with wavelengths $>r_c$) and the low-energy modes controllable via our feedback. In principle, this method could even work for systems with highly non-trivial three-dimensional dynamics, provided thermalisation is rapid enough to allow energy extraction purely from controlling excitations visible in the measured column density. Furthermore, while our simple control loop based on feedback of the derivative signal is highly effective, it is not necessarily optimal. The results presented here may be further improved with sophisticated optimization techniques to produce superior control loops. With the application of more sophisticated control loops, feedback cooling of Bose gases may also be used in more exotic scenarios, such as to create dipolar or molecular condensates. Finally, there may be great benefit to adapting the methods presented here to the cooling and control of ultracold Fermi gases, potentially aiding in the realization of novel states of quantum matter.

\newpage
\section{Methods}
\subsection{Reduction to 1D}
To derive an effective-1D description of the Bose gas, we consider a harmonic trap in which the axial confinement is stronger than the longitudinal confinement ($\omega_\perp \gg \omega_x$). In this limit, the transverse dynamics are strongly suppressed, allowing the following ansatz for the full quantum field $\hat{\psi}(\bm{r})$
\begin{equation}
    \hat{\psi}(x,\bm{r}_\perp,t)=\hat{\Psi}(x,t)\phi(\mathbf{r}_\perp)
\end{equation}
where $\bm{r}_\perp=(y,z)$. A reasonable choice for the radial profile $\phi$ is a Gaussian of standard deviation $R_\perp$
\begin{equation}
    \phi(\bm{r}_\perp) =\frac{1}{\sqrt{\pi}R_\perp}\exp\left(-\frac{|\bm{r}_\perp|^2}{2R_\perp^2}\right) \,.
\end{equation}
Integrating over the axial profile gives the effective 1D interaction strength
\begin{equation}
    U_\text{1D}=\frac{2\hbar^2 a_s}{m R_\perp^2} \,,
\end{equation}
where $a_s \approx 100a_0$ and $m$ are the $s$-wave scattering length and atomic mass, respectively, for ${}^{87}\text{Rb}$. We choose $R_\perp$ such that the Gross-Pitaevskii energy functional is minimised for the axial Gaussian ansatz. In the Thomas-Fermi limit, this gives [Equation (6.27) in Ref.~\cite{Pethick2008}]:
\begin{equation}
\label{eq:VariationalGasWidth}
    R_\perp=\left(\frac{2}{\pi}\right)^{1/10}\left(\frac{N a_s}{\bar{a}}\right)^{1/5}\frac{\bar{\omega}}{\omega_\perp}\bar{a}
\end{equation}
where $\bar{a}=\sqrt{\hbar/m\bar{\omega}}$ and $\bar{\omega}=(\omega_x\omega_\perp^2)^{1/3}$. For the experimental parameters considered, we find the above estimate to be within $6\%$ of the true variational minimum determined by numerically solving Equation~(6.22) in Ref.~\cite{Pethick2008}. For $N=10^4$ ${}^{87}$Rb atoms with $(\omega_x,\omega_\perp)=2\pi\times(40,400)$ Hz, the above expression gives $R_\perp\approx 0.477 x_0$, where $x_0=\sqrt{\hbar/m\omega_x}$. This corresponds to an effective interaction strength of $U_{1\text{D}}/x_0\approx 0.027\hbar\omega_x$. \par

The conditional master equation (\ref{eqn:conditionalmasterequation}) is given already in the effective-1D regime, and further details are provided in Ref.~\cite{Szigeti2009}.

\subsection{Measurement parameterisation}
For the measurement operator $\hat{M}(x)\equiv \int dx'\hat{\Psi}^\dagger(x')K(x-x')\hat{\Psi}(x')$, the point-spread kernel for the optical diffraction limit is conveniently expressed in $k$-space:
\begin{align}
    K(x)&\equiv\frac{1}{\sqrt{2\pi}}\int dk \tilde{K}(k)e^{ikx} \nonumber\\
    \tilde{K}(k)&\equiv\sqrt{\frac{r_d}{2\Gamma(5/4)}}e^{-(r_dk)^4/2}
\end{align}
where
\begin{equation}
r_d\equiv\frac{\sqrt{\lambda R_\perp/\pi}}{2^{3/4}}
\end{equation}
is the resolution scale imposed by the optical diffraction limit, and $\Gamma(x)$ is the gamma function. This parameterisation of $K(x)$ differs from that of Refs.~\cite{Szigeti2009,Szigeti2010,Hush2013} in that $K(x)$ is normalised to unity, thereby ensuring that the measurement strength $\alpha$ and optical diffraction limit $r_d$ can be varied independently (Plancherel's Theorem). Consequently, the measurement strength $\alpha$ is:
\begin{equation}
\label{eq:alpha}
    \alpha \equiv \frac{3\Gamma(5/4)}{2^{21/4}\pi^4}\left(\frac{\lambda}{R_\perp}\right)^{3/2}\frac{\Omega^2}{\Delta^2}\Delta\nu
\end{equation}
where $\Omega$, $\Delta$ and $\Delta\nu$ are the Rabi frequency, detuning from the ${\rm D}_2$ transition ($5^2{\rm S}_{1/2}\to 5^2{\rm P}_{3/2}$), and excited state linewidth, respectively.

\subsection{Imperfections in control and measurement}
Previous works considered the use of a simple control that manipulated the width and position of a harmonic trap to damp the sloshing and breathing modes of a BEC \cite{Haine2004,Szigeti2009,Szigeti2010}, with higher-order modes coupled to the control via the nonlinear atom-atom interactions. A `quantum-noise control' for the cosine-squared mode was also considered in Ref.~\cite{Hush2013} to counteract beyond-mean-field heating channels. However, the more complicated thermal excitations here necessitate a more sophisticated control. Recent innovations in high-speed configurable optical potentials \cite{Gauthier2016} allow us to now consider experimentally achievable controls of far greater spatial complexity. In this work we use a feedback inspired by the energy-damping thermal dissipation process in the semiclassical thermal reservoir theory for the Bose gas~\cite{McDonald2020} (often referred to as the \emph{scattering} process in earlier works~\cite{Blakie2008}):
\begin{align}
    V_{\rm C}^\text{ideal}(x,t)&=k_{\text{ED}}\partial_t\rho(x,t),\\
    \rho(x,t)&=\langle \hat{\Psi}^\dagger(x) \hat{\Psi}(x) \rangle_c,
\end{align}
which directly opposes the motion of the atomic cloud and counteracts spatial excitations at all orders.  $k_{\text{ED}}$ is a gain factor for the control and chosen to be the near-optimal value $k_{\text{ED}}=15.0 / N$. However, the experimenter will not have perfect knowledge of $\rho(x,t)$, nor will they have perfect control of the feedback potential due to the resolution limit of the SLM set by the optical diffraction limit. While the effects of imperfect knowledge are often modelled via a Bayesian estimation framework for systems of this type \cite{Szigeti2013}, we instead consider an experimentally-relevant scenario in which the experimenter directly feeds back the derivative signal. Specifically, the continuous measurement results in a measurement signal $Y(x,t)=2\sqrt{\eta\bar{\alpha}}\tilde{\rho}(x,t)+\Theta(x,t)$, where $\tilde{\rho}(x,t)\equiv\int dx' K(x-x')\rho(x',t)$ is the diffraction-limited density, and $\Theta(x,t)$ is corrupting Gaussian white noise with zero mean and unit variance. This corrupting white noise arises due to the fundamental incompatibility of the system observables associated with our measurement. In practice, the experimenter constructs the derivative signal by finite differencing over a small time interval $\Delta t$:
\begin{equation}
    V_{\rm C}(x,t)\equiv \frac{k_{\text{ED}}}{2\sqrt{\eta\bar{\alpha}}}\int dx' G_{r_c}(x-x')\frac{Y(x',t)-Y(x',t-\Delta t)}{\Delta t}
\end{equation}
where we have additionally convolved the control with a Gaussian point-spread function $G_{r_c}(x)=\frac{1}{\sqrt{2\pi}}\int dk e^{ikx}\tilde{G}_{r_c}(k)$ with full-width at half-maximum of the SLM resolution $r_c$. This models the finite spatial resolution of the control. Due to the high switching speed of available SLMs \cite{Gauthier2016}, we may take the limit $\Delta t \to 0$, yielding:
\begin{equation}
    V_{\rm C}(x,t)=k_{\text{ED}}\int dx' G_{r_c}(x-x')\left(\partial_t \tilde{\rho}(x',t)+\frac{1}{2\sqrt{\eta\bar{\alpha}}}\zeta(x',t)\right),
    \label{eqn:controlpotential}
\end{equation}
where $\zeta(x,t)$ is an infinitesimal Stratonovich noise increment; it directly corresponds to the quantum measurement noise described by the It\^{o} increment $dW(x,t)$ in Eq.~\ref{eqn:conditionalmasterequation}.
\subsection{Number-phase Wigner unravelling}
The master equation (\ref{eqn:conditionalmasterequation}) can be recast in Stratonovich form:
\begin{equation}
    d\hat{\rho}_c=-i[\hat{H},\hat{\rho}_c]dt+\bar{\alpha}\int dx \mathcal{D}[\hat{M}(x)]\hat{\rho}_c dt + \eta\bar{\alpha} \int dx \mathcal{C}[\hat{M}(x)]\hat{\rho}_c+ \sqrt{\eta\bar{\alpha}}\int dx \, \mathcal{H}[\hat{M}(x)]\hat{\rho}_c \zeta(x,t),
    \label{eqn:stratonovichSME}
\end{equation}
where $\mathcal{C}[\hat{L}]\hat{\rho}\equiv \langle\hat{L}+\hat{L}^\dagger\rangle\mathcal{H}[\hat{L}]\hat{\rho}-\frac{1}{2}\mathcal{H}[\hat{L}^2]\hat{\rho}+\langle\hat{L}^\dagger\hat{L}\rangle\hat{\rho}-\hat{L}\hat{\rho}\hat{L}^\dagger$ is the Stratonovich superoperator. We have written Eq.~(\ref{eqn:stratonovichSME}) in harmonic oscillator units, where distance, time, and energy are in units of $x_0=\sqrt{\hbar/m\omega_x}$, $\omega_x^{-1}$, and $\hbar\omega_x$, respectively, and $\bar{\alpha}\equiv \alpha/\omega_x$ is the dimensionless measurement strength.\par

Equation~(\ref{eqn:stratonovichSME}) can be tractably simulated by exactly mapping the conditional evolution to the number-phase Wigner (NPW) representation and, after a series of approximations valid in the large-occupation-per mode-regime, applying a stochastic unravelling of the dynamics via the NPW particle filter~\cite{Hush2013}. The conditional density matrix $\hat{\rho}_c$ is encoded in a swarm of fields $\psi^{(j)}(x,t)$ and corresponding weights $W^{(j)}(t)$, where the upper index $(j)$ corresponds to a `fictitious' stochastic unravelling (analogous to the stochastic unravelling of the truncated Wigner method \cite{Blakie2008}). The fields and weights evolve as
\begin{subequations}
\label{eqn:NPWxspace}
\begin{align}
    \partial_t \psi^{(j)}(x,t)&=-i\left(h^{(j)}_{\text{int}}(x,t,U)\psi^{(j)}(x,t)+\sqrt{\bar{\alpha}}\int dx' K(x-x',r)\psi^j(x,t)\xi^{(j)}(x',t)\right)\label{eqn:xspacefields} \\
    \partial_t W^{(j)}(t)&=2W^{(j)}(t)\left(C^{(j)}(t)+I^{(j)}(t)\right) \label{eqn:weights}
\end{align}
\end{subequations}
where $h^{(j)}_{\text{int}}(x,t,U_{1\text{D}})\equiv -\frac{1}{2}\partial_x^2+\frac{1}{2}x^2+\bar{V}_C(x,t)+U_{1D}|\psi^{(j)}(x)|^2$ is an effective Hamiltonian for each field sample $(j)$, $\bar{V}_C(x,t)\equiv V_{\rm C}(x,t)/\hbar\omega_x$ is the control potential in harmonic oscillator units, $\xi^{(j)}(x',t)$ is a Stratonovich noise increment originating from the fictitious stochastic unravelling, and
\begin{subequations}
\begin{align}
C^{(j)}(t)&\equiv \int dx M^{(j)}(x,t)\left(2\mathbb{W}\left[M^{(\cdot)}(x,t)\right]-M^{(j)}(x,t)\right)\\
I^{(j)}(t)&\equiv \int dx M^{(j)}(x,t) \zeta(x,t)\\
M^{(j)}(x,t)&\equiv \sqrt{\eta\bar{\alpha}}\int dx' K(x-x')\rho^{(j)}(x',t)
\end{align}
\end{subequations}
for the spatial density sample $\rho^{(j)}(x,t)\equiv |\psi^{(j)}(x,t)|^2-\frac{1}{2}\delta(x,x)$. The $\delta(x,x)$ term is the \textit{Wigner correction}; it is formally infinite on the diagonal, but has a finite projection onto the bases considered here and so can be self-consistently included in our simulation. Conditional expectation values correspond to weighted averages over fictitious paths:
\begin{equation}
    \langle:\hat{\Psi}^\dagger(x, t)\hat{\Psi}^\dagger(y, t)\dots\hat{\Psi}(z, t):_{\text{sym}}\rangle_c=\mathbb{W}\left[{\psi^{(\cdot)^*}}(x,t){\psi^{(\cdot)^*}}(y,t)\dots\psi^{(\cdot)}(z,t)\right]
    \label{eqn:expectationvalues}
\end{equation}
where $:\hat{A}\hat{B}\dots \hat{C}:_{\text{sym}}$ is the symmetric ordering of the field operator product $\hat{A}\hat{B}\dots \hat{C}$ and 
\begin{equation}
    \mathbb{W}\left[f^{(\cdot)}\right]\equiv\frac{\sum_j W^{(j)}(t)f^{(j)}}{\sum_k W^{(k)}(t)}
\end{equation}
defines the weighted stochastic average. Equation \ref{eqn:expectationvalues} is formally exact in the limit of infinite fictitious trajectories. Equation \ref{eqn:expectationvalues} is formally exact in the limit of infinite fictitious trajectories. We find that 100-300 fictitious trajectories are sufficient for good convergence of our simulations.\par

It is crucial to note that Eq.~ \ref{eqn:conditionalmasterequation} and its NPW unravelling in Eq.~\ref{eqn:NPWxspace} describe the evolution of the \textit{conditional} density matrix $\hat{\rho}_c$, conditioned on a particular record of stochastic quantum measurement outcomes $\zeta(x,t)$. To describe the average behaviour of an initially thermal cloud, and thus compute ensemble values such as condensate fraction, a second round of averaging is required (over `real' trajectories corresponding to different measurement records $\zeta(x,t)$). We simulate the projective form of Equation \ref{eqn:NPWxspace} multiple times with different initial samples and different random measurement records $\zeta(x,t)$, compute conditional expectation values (\ref{eqn:expectationvalues}), then take an unweighted mean over all real trajectories. It should be noted that fictitious trajectories within a given real trajectory are coupled to each other via the weights (\ref{eqn:weights}), but real trajectories are not coupled to each other (as they represent incoherent samples of a mixed state). The number of real trajectories required for acceptable confidence intervals (computed by a bootstrapping procedure) vary with temperature and calculated quantity, but we find 100-200 real trajectories to be more than sufficient for the regimes and quantities studied here.\par

This is a full-field method, and all terms in the unravelled NPW evolution (\ref{eqn:NPWxspace}) are exact compared to the conditional master equation (\ref{eqn:conditionalmasterequation}), except for the kinetic energy term, which takes an approximate form valid in the large atom number regime. It was verified in Ref.~\cite{Hush2013} that $N=10^2$ atoms is sufficient for excellent agreement of the kinetic energy term with an exact solution. Here we typically consider $N=10^4$ and no less than $N=3\times 10^3$ atoms, which clearly is sufficient. An additional requirement is that our initial state needs to be well-represented by a non-negative phase-space distribution - which is certainly true for the incoherent thermal states considered here.

\subsection{Weights and resampling}
Rather than directly evolving the weights by Eq.~\ref{eqn:weights}, we evolve the weights in log-space ($w^{(j)}\equiv \log(W^{(j)})$) for improved numerical stability:
\begin{equation}
\partial_t w^{(j)}(t)=\left(C^{(j)}(t)+I^{(j)}(t)\right).
\end{equation}
We also renormalise the weights to $\sum_j W^{(j)}=1$ when their norm exceeds a certain threshold to maintain numerical stability. This is more stable than modifying Eq.~\ref{eqn:weights} to preserve the norm, as it avoids loss of significance from difference terms at each step. Naively integrating this will rapidly result in a single fictitious path dominating the ensemble ($W^{(j)}\approxeq 1$ for some weight and $W^{(k)}\approxeq 0$ for all others). The NPW unravelling (\ref{eqn:NPWxspace}) is a particle filter for the conditional master equation (\ref{eqn:conditionalmasterequation}), and this phenomenon is known as \textit{sample impoverishment} in particle filter literature. This can be mitigated by regularly resampling the field - replacing negligibly-weighted paths with copies of highly-weighted paths, appropriately re-weighting the fields, and subsequently evolving them under different fictitious noises $\xi^{(j)}(x,t)$. In Ref.~\cite{Hush2013} this was achieved with a naive `breeding' algorithm, but it was noted that it was not the optimal solution to the resampling problem.\par

Here we achieve superior stability in NPW by using sequential importance resampling (SIR), a standard resampling technique for particle filters. This added stability assists the simulation of more challenging initial states, such as low-condensate-fraction thermal clouds. At each time step we calculate the effective number of samples $N_s\in [1,K]$
\begin{equation}
    N_s \equiv \frac{(\sum_j W_j)^2}{\sum_k W_k^2},
\end{equation}
where $K$ is the number of fictitious paths. If $N_s/K < c$, where $c\in(0,1)$ is some threshold (typically chosen to be $c=\frac{1}{2}$), then we resample all fields $\psi^{(j)}(x,t)$ according to the categorical distribution formed by their weights $W^{(j)}$, and subsequently reset all weights to $W^{(j)}=1/K$. In practice, this is achieved efficiently using the sorted-deterministic algorithm from Ref.~\cite{Kitagawa1996}.

\subsection{Self-consistent projection of NPW dynamics}
To correctly model thermal phenomena, we combine the NPW particle filter with projective methods from c-field theory. An excellent introduction to the motivations and techniques of c-field theory is provided in Ref.~\cite{Blakie2008}. In essence, the quantum field theory must be regularised by imposing an initial-temperature-dependent energy cutoff, which is achieved by projecting the system's dynamics onto a finite basis of non-negligibly occupied modes. However, the natural modes of a spatial grid representation (i.e. plane wave modes) do not correspond to the eigenmodes of a harmonically-trapped system, meaning that the position-space NPW particle filter derived in Ref.~\cite{Hush2013} cannot describe thermal phenomena self-consistently on a finite spatial grid.\par

To overcome this limitation, we instead consider evolution self-consistently projected onto non-negligibly occupied coherent modes of the trap
\begin{equation}
    c_n^{(j)}(t)\equiv \mathcal{P}\left\{\int dx \phi_n(x) \psi^{(j)}(x,t)\right\},
\end{equation}
where $\phi_n(x)$ is the $n$th eigenmode of the harmonic oscillator, and $\mathcal{P}$ is a projector onto the coherent region such that $\mathcal{P}\left\{\psi^{(j)}(x,t)\right\} = \sum_n^{n_{\text{cut}}} c_n^{(j)}(t)\phi_n(x)$. We choose our cutoff mode $n_{\text{cut}}$ for the coherent region to be sufficiently large such that our results do not change with increasing basis size, which for the temperatures simulated is no more than $n_{\text{cut}}=100$.\par

We construct the equations of motion for the fields $c_n^{(j)}(t)$ by projecting Eq.~\ref{eqn:xspacefields} as $\partial_t c_n^{(j)}(t)=\mathcal{P}\left\{\int dx \phi_n(x) \partial_t \psi^{(j)}(x,t)\right\}$, and recasting Equation \ref{eqn:weights} in terms of $c_n^{(j)}(t)$. We make extensive use of Hermite-Gaussian quadratures in computing $x$- and $k$-space integrals, allowing all terms to be computed either exactly to machine precision (where point-spread kernels are not involved) or extremely precisely (for terms involving point-spread kernels). We note that most non-diagonal terms in the evolution are weighted sums of overlap integrals
\begin{equation}
    \mathcal{I}=\int dx \prod_m \phi_m(x),
\end{equation}
which can be computed exactly using Hermite-Gaussian quadrature weights and points \cite{Abramowitz1965}. We refer to an integral over a product of $N$ eigenmodes $\phi_n(x)$ as an $N$\textit{-field quadrature}. The necessary expressions can be obtained by writing convolutions in $k$-space, computing Fourier transforms via the harmonic oscillator eigenbasis, and exploiting mathematical properties of the eigenmodes $\phi_n(x)$. The weights evolve by Eq.~\ref{eqn:weights}, where:
\begin{align}
    C^{(j)}(t)&=2\pi\eta\bar{\alpha}\left(\left(\sum_{nm}^{n_{\text{cut}}}2\rho_n^{(j)}(t)\mathbb{W}\left[\rho_m^{(\cdot)}(t)\right]g_{nm}\right)-\left(\sum_{nm}^{n_{\text{cut}}}\rho_n^{(j)}(t)\rho_m^{(j)}(t)g_{nm}\right)\right),\label{eqn:cjHGbasis} \\
    I^{(j)}(t)&=\sqrt{2\pi\eta\bar{\alpha}}\int dk \mathcal{P}\left\{\tilde{K}(k)\right\}\left(\sum_m^{n_{\text{cut}}} (-i)^m \rho_m^{(j)}(t)\phi_m(k)\right)\left(\sum_n^{n_{\text{cut}}} i^n \zeta_n(t)\phi_n(k)\right),\\
    g_{nm}&\equiv i^{m-n}\underbrace{\int dk \phi_n(k) \left(\mathcal{P}\left\{\tilde{K}(k)\right\}\right)^2 \phi_m(k)}_{\text{4-field quadrature}},\\
    \rho_{n}^{(j)}&\equiv \mathcal{P}\left\{\rho^{(j)}(x,t)\right\}=\underbrace{\int dx \phi_n(x)\left(\left|\mathcal{P}\left\{\psi^{(j)}(x,t)\right\}\right|^2-\frac{1}{2}\sum_m \phi_m(x)\phi_m(x)\right)}_{\text{3-field quadrature}},\label{eqn:projecteddensity} \\
\end{align}
and $\zeta_n(t)\equiv \mathcal{P}\left\{\int dx \phi_n(x) \zeta(x,t)\right\}$ are Stratonovich measurement noise processes that may be generated independently without reference to $x$-space. We similarly generate moments of the fictitious noise $\xi_n(t)\equiv \mathcal{P}\left\{\int dx \phi_n(x) \xi(x,t)\right\}$ directly in the Hermite-Gauss basis. We note that the formally infinite Wigner correction has a finite projection in the second term of Equation \ref{eqn:projecteddensity}. It is crucial to compute Equation \ref{eqn:cjHGbasis} in the bracketed order; if the sum is computed after the difference, then this results in numerical instability due to accumulation of many loss-of-significance errors.\par

The field evolution can be computed as:
\begin{align}
\partial_t c_n^{(j)}(t) =& -i\left(n+\frac{1}{2}\right)c_n^{(j)}(t)-iU_{1D}\overbrace{\int dx \phi_n(x)\left|\mathcal{P}\left\{\psi^{(j)}(x,t)\right\}\right|^2\mathcal{P}\left\{\psi^{(j)}(x,t)\right\}}^{\text{4-field quadrature}}\nonumber\\&-i\underbrace{\mathcal{P}\left\{\int dx \phi_n(x)V_{\rm C}(x,t)\psi^{(j)}(x,t)\right\}}_{\text{3-field quadrature}}-i\sqrt{2\pi\bar{\alpha}}\sum_m^{n_{\text{cut}}} \xi^{(j)}_m(t)\underbrace{\int dx \phi_n(x)\mathcal{P}\left\{\psi^{(j)}(x,t)\right\}(x,t)f_m(x)}_{\text{3-field quadrature}}
\end{align}
where $f_{nm}\equiv i^{n-m}\int dk \phi_n(k)\mathcal{P}\left\{\tilde{K}_r(k)\right\}\phi_m(k)$ is pre-computed with a 3-field quadrature. The control term (for the derivative signal feedback defined in Equation \ref{eqn:controlpotential}) is projected as:
\begin{align}
    \mathcal{P}\left\{\int dx \phi_n(x)V_{\rm C}(x,t)\psi^{(j)}(x,t)\right\}=&k_{\text{ED}}\Bigg(\sum_m^{n_{\text{cut}}} \tilde{J}_m(t)\underbrace{\int dx \phi_n(x)\phi_m(x)\mathcal{P}\left\{\psi^{(j)}(x,t)\right\}}_{\text{3-field quadrature}}\nonumber\\&+\frac{1}{2\sqrt{\eta\bar{\alpha}}}\sum_m^{n_{\text{cut}}} \tilde{\zeta}_m(t)\underbrace{\int dx \phi_n(x)\phi_m(x)\mathcal{P}\left\{\psi^{(j)}(x,t)\right\}}_{\text{3-field quadrature}}\Bigg), \\
    \tilde{\zeta}_n(t)\equiv&\sum_m^{n_{\text{cut}}}i^{n-m}\zeta_n(t)\underbrace{\int dk \tilde{G}_{r_c}(k)\phi_n(k)\phi_m(k)}_{\text{3-field quadrature}}, \\
    \tilde{J}_n(t)\equiv&\sum_n^{n_{\text{cut}}} i^{m-n}J_n(t)\underbrace{\int dk \tilde{G}_{r_c}(k)\tilde{K}(k)\phi_m(k)\phi_n(k)}_{\text{4-field quadrature}},
\end{align}
where $J_n(t)\equiv \mathcal{P}\left\{\int dx \phi_n(x)\partial_t\langle\hat{\Psi}^\dagger(x)\hat{\Psi}(x)\rangle_c \right\}$ is the projection of the local rate of change in column density. By applying the continuity theorem $\partial_t\langle\hat{\Psi}^\dagger(x)\hat{\Psi}(x)\rangle_c=-\text{Im}\lbrack\langle\hat{\Psi}^\dagger(x)\partial_x^2\hat{\Psi}(x)\rangle_c\rbrack$  and applying Equation \ref{eqn:expectationvalues}, this may be written as
\begin{equation}
    J_n(t)=-\mathbb{W}\left[\underbrace{\int dx \phi_n(x)\mathcal{P}\left\{\psi^{(\cdot)^*}(x,t)\right\}\mathcal{P}\left\{\partial_x^2 \psi^{(\cdot)}(x,t)\right\}}_{\text{4-field quadrature}}\right].
\end{equation}
By expanding $\partial_x$ into ladder operators on the eigenmodes $\phi_n(x)$, the second derivative term may be written as
\begin{equation}
    \mathcal{P}\left\{\partial_x^2 \psi^{(j)}(x,t)\right\}=\mathcal{P}\left\{\sum_n c_n^{(j)}(t)\left(\sqrt{n}\sqrt{n-1}\phi_{n-2}(x)-(2n+1)\phi_n(x)+\sqrt{n+1}\sqrt{n+2}\phi_{n+2}(x)\right)\right\}
\end{equation}
where the projector $\mathcal{P}$ acting on the sum means that we set $\phi_n(x)\to0$ for $n\notin[0,n_{\text{cut}}]$. \par

The measurement kernel $\tilde{K}(k)$ and control kernel $\tilde{G}_{r_c}(k)$ are well-represented by Hermite-Gauss modes, and all other terms are exact within the self-consistent projection of the basis. Consequently, this marriage of projective c-field methods and the NPW particle filter enables us to efficiently and self-consistently simulate the dynamics of an initially thermal Bose gas to near-exact precision, so long as we choose an appropriate cuttof $n_{\text{cut}}$. All dynamical equations simulated in this work were integrated using XMDS2 \cite{Dennis2013}, which allows exact computation of $N$-field quadrature integrals and supports a variety of integration algorithms. Since the dominant contributions to the integration timescale are deterministic, we find that an adaptive fourth-fifth order Runge-Kutta algorithm yields good convergence for most simulations, but also use fixed-step fourth-order Runge-Kutta with a smaller timestep for simulations including feedback lag, simulations with more than $10^4$ atoms, and validation of convergence for select other simulations.
\subsection{Condensate fraction \& sampling thermal states}
We use the Penrose-Onsager definition to compute condensate fraction \cite{Penrose1956}. We compute the one-body density matrix
\begin{equation}
    G^{\text{1B}}_{nm}(t)=\langle \hat{c}^\dagger_n \hat{c}_m \rangle=\mathbb{E}[\langle \hat{c}^\dagger_n \hat{c}_m \rangle_c]=\mathbb{E}[\mathbb{W}[c_n^{(\cdot)^*}c_m^{(\cdot)}-\tfrac{1}{2}\delta_{nm}]]
\end{equation}
where $\hat{c}_n^\dagger\equiv \int dx \phi_n(x) \hat{\Psi}(x)$ are field operators in the trap basis, and $\mathbb{E}\left[\cdot\right]$ is an average over real trajectories. We diagonalise $G^{\text{1B}}_{nm}(t)$ to find its eigenvalues $\mathcal{G}_n^{\text{1B}}(t)$. At a given time $t$, the condensate fraction $f_{\text{condensate}}$ is given by
\begin{equation}
    f_{\text{condensate}}\equiv\frac{\max_n\left(\mathcal{G}_n^{\text{1B}}(t)\right)}{\sum_m\mathcal{G}_m^{\text{1B}}(t)}
\end{equation}
It is much more efficient to compute the one-body density matrix in the trap basis \cite{Blakie2008}. Computing it in $x$-space is technically possible, but significantly increases sampling requirements for accurate results.\par

The initial states of our simulations are thermal states in the grand canonical ensemble:
\begin{equation}
    \hat{\rho}=\frac{\exp(-\beta\hat{H}_{GC})}{\Tr\left\{\exp(-\beta\hat{H}_{GC})\right\}},
\end{equation}
where $\beta\equiv \hbar\omega_x/k_BT$ and $\hat{H}_{GC}\equiv \hat{H} - \mu\sum_n \hat{c}^\dagger_n\hat{c}_n$. Although it is possible to directly sample this full mixed state in NPW, this leads to numerical instabilities in the weight evolution and rapid sample impoverishment. The NPW particle filter is numerically stable and well-tested for coherent states \cite{Hush2012,Hush2013}, so we instead note that this state can be written as a mixture of coherent states:
\begin{equation}
    \hat{\rho}=\int d^2\bm{\alpha} P(\bm{\alpha}) \ket{\bm{\alpha}}\bra{\bm{\alpha}}
\end{equation}
and reproduce the correct average statistics by conditionally initialising the NPW particle filter in a pure coherent state, with order parameter $\bm{\alpha}$ randomly sampled from $P(\bm{\alpha})$. This sampling is performed using the simple-growth stochastic projected GPE (SPGPE), which evolves any initial state to samples of the grand-canonical ensemble \cite{Blakie2008}. We choose the temperature $T$ and the chemical potential $\mu$ in the SPGPE such that the ensemble average of the samples have the desired initial condensate fraction and number of atoms. We then conditionally initialise the NPW filter in the sampled state using the coherent-state sampling algorithm from Ref.~\cite{Hush2012}. This approach achieves superior convergence by efficiently dividing samples of the thermal distribution between `real' and `fictitious' trajectories in a way that maximizes numerical stability of the particle filter.\par

We find that a basis of $100$ Hermite-Gauss modes is sufficient to represent the field in the regimes investigated here, even for the largest simulated values of initial temperature. Further increasing the basis size at the temperatures simulated has no significant quantitative impact on relevant observables such as condensate fraction.

\subsection{Integrated signal-to-noise measure \label{method:IntegratedSNR}}
The information contained within an optical measurement of the atomic density is fundamentally connected to the amount of spontaneous emission induced by the light field, i.e. the destructiveness of the measurement. For a shot-noise limited detection of the light with some inefficiency $\eta$, the SNR of the two-dimensional column density image is given by \cite{Hope2004,Hope2005}
\begin{align}
    {\rm SNR} &= \frac{\bar{n}}{2}\sqrt{\eta A\sigma \Gamma_{\rm sp}\Delta t_\text{d}} \,,
\end{align}
where $\Gamma_{\rm sp}$ is the spontaneous emission rate and $\Delta t_\text{d}$ is the duration of the measurement. Squaring both sides, taking the infinitesimal limit $\Delta t_\text{d}\rightarrow dt$, and integrating over a time window $t\in [0,\tau]$ then gives:
\begin{align}
    \text{SNR}_\text{I}^2 \equiv \int \text{SNR}^2 &= \left(\frac{\bar{n}}{2}\right)^2\eta A\sigma \Gamma_{\rm sp} \tau \,.
\end{align}
This expression defines the \emph{integrated signal-to-noise} $\text{SNR}_\text{I}$ for a continuous measurement over time $\tau$. The strength of the measurement is implicit in the spontaneous emission rate $\Gamma_{\rm sp}$, which we have assumed to be constant here for simplicity (corresponding to fixed power in the light field). The total fraction of atoms lost during time $\tau$ can then be expressed in terms of the integrated SNR as:
\begin{align}
    \frac{N_{\rm lost}}{N_0} &=1- e^{-\Gamma_\text{sp} \tau} = 1- \exp\left(-\frac{4\;\text{SNR}_\text{I}^2}{\bar{n}^2 \eta A \sigma} \right) \,.
\end{align}
To calculate the fraction of atoms lost for a given $\text{SNR}_\text{I}$, we can estimate the averaged column density on a given pixel as $\bar{n} \approx N/(2\pi R_x R_y)$ with the gas widths $R_i$ given by the Gaussian variational estimate Eq.~\eqref{eq:VariationalGasWidth}. These expressions give that the gas width increases with number due to interatomic interactions as $R {\sim} N^{1/5}$, leading to  $\bar{n}{\sim}N^{3/5}$. Substituting this scaling into the expression above and Taylor expanding to first order (assuming large $N$) then gives $N_\text{lost}/N_0 \sim N^{-6/5}$. \par

In this work we consider a feedback signal based on the \emph{one-dimensional} column density, which is constructed by integrating over the tightly-trapped spatial dimension. Taking the noise on each pixel to be uncorrelated, this one-dimensional column density will have an SNR a factor of $\sqrt{m}$ larger than the two-dimensional image, where $m$ is the average number of pixels covering the condensate along the tightly-trapped dimension. We may crudely estimate $m$ based on the ratio of the ground-state width of the gas in the transverse dimension $R_\perp\sim N^{1/5}$ to the resolution of the DMD $r_c$, i.e. $m\approx R_\perp/r_c$. If we assume $r_c$ is roughly independent of number, then we find $N_\text{lost}/N_0 \sim N^{-7/5}$ in the limit of large $N$.

\subsection{Estimation of atomic loss due to spontaneous emission}
In order to estimate the atomic loss due to heating induced by the imaging laser, we treat each spontaneous emission event as resulting in the loss of an atom from the trap. This provides a conservative estimate of the fraction of atoms remaining after feedback. The rate of spontaneous emission can be calculated as the product of the excited state population $P_e = \frac{\Omega^2}{4\Delta^2}$ and the linewidth of the excited state $\Delta \nu$ \cite{Hope2004}, which is proportional to the measurement strength $\alpha$ defined in Eq.~\eqref{eq:alpha}:
\begin{equation}
\label{eq:spontymitty}
    \Gamma_\text{sp}\equiv\Delta\nu =\alpha \frac{2^{13/4}\pi^4}{3\Gamma(5/4)}\left(\frac{R_\perp}{\lambda}\right)^{3/2} \,.
\end{equation}
The total number of atoms remaining after a measurement period of $\tau$ is then simply calculated as $N(t) = N(0)\exp(-\Gamma_\text{sp}\tau)$.

\bibliographystyle{bibsty}
\bibliography{bibliography}
\section{Acknowledgements}
The authors acknowledge fruitful discussions with Russell Anderson, Ethan Barden, John Close, Matthew Davis, Simon Haine, Robert Nyman, and Nicholas Robins. This project was supported by Australian Research Council (ARC) Discovery Project No.~DP190101709. SSS received funding from an Australia Awards-Endeavour Research Fellowship and an ARC Discovery Early Career Researcher Award (DECRA), project No. DE200100445. MLG was supported by a Rhodes Scholarship. ZM and RLT were supported by an Australian Government Research Training Program (RTP) Scholarship. ASB acknowledges support from the Marsden Fund (Grant No. UOO1726) and the Dodd-Walls Centre for Photonic and Quantum Technologies. This research was undertaken with the assistance of resources and services from the National Computational Infrastructure (NCI), which is supported by the Australian Government.
\end{document}